\documentclass{ws-jmrr}
\usepackage[sort,compress,super]{cite}

\usepackage{graphics} 
\usepackage{graphicx}
\usepackage{mathptmx} 
\usepackage{times} 
\usepackage{amsmath} 
\usepackage{amssymb}  
\usepackage{array}
\usepackage{color}
\usepackage[table]{xcolor}
\usepackage{bm}
\usepackage{soul}
\usepackage{array,multirow}
\usepackage{array}
\usepackage{makecell}
\usepackage{booktabs}
\usepackage{float}

\begin{document}

\catchline{0}{0}{2018}{}{}
\markboth{Florence Leong}{Dynamics of Insufflated Abdominal Wall Tissue for Magnetically Anchored Surgical Instruments}

\title{Dynamics of Insufflated Abdominal Wall Tissue for Magnetically Anchored Surgical Instruments}

\author{Florence Leong$^{a,}$\footnote{Melbourne School of Engineering, University of Melbourne, Parkville VIC 3010, Australia}, Alireza Mohammadi$^{a}$, Vijay Rajagopal$^{a}$, Ying Tan$^{a}$, Dhan Thiruchelvam$^{b}$, \\ Pietro Valdastri$^{c}$ and Denny Oetomo$^{a}$}

\address{$^a$Melbourne School of Engineering, University of Melbourne, Parkville VIC 3010, Australia\\
E-mail: florence.leong@unimelb.edu.au}

\address{$^b$Department of Surgery, University of Melbourne, St Vincent's Hospital, 
Fitzroy, VIC 3065, Australia\footnote{Department of Surgery, University of Melbourne at St Vincent's Hospital, Fitzroy, VIC 3065, Australia}}

\address{$^c$School of Electronic and Electrical Engineering, University of Leeds, Leeds LS2 9JT, UK\footnote{School of Electronic and Electrical Engineering, University of Leeds, Leeds LS2 9JT, UK}}



\maketitle

\begin{abstract}

Magnetically-anchored surgical devices have recently gained attention in abdominal surgery applications, with the implementation of magnets to hold or anchor surgical devices onto the insufflated abdominal wall (MAGS). These anchors have been used to secure passive and active devices, where active device such as robotic manipulators, produce motions that would excite the dynamics of the non-rigid abdominal walls. Hence, there is a need to investigate the mechanical dynamics of the abdominal wall tissue in insufflated state, combined with the magnetic anchoring, specifically its response to mechanical excitations and the expected resulting disturbances to the operation of the anchored devices. In this paper, loading and unloading tests are performed on a corresponding porcine specimen for dynamics identification. The experimental setup was constructed to emulate the insufflated state of the abdomen with the magnetically anchored mechanism. The tissue creep responses during unloading are captured and approximated with a general numerical model, which is in turn used for the dynamic analysis of the abdominal wall tissue using Bode plot. The results  showed that in such stretched and compressed state, the steady state displacement of the abdominal wall is zero, with a time constant of 0.2 secs, 0.085 secs and 0.045 secs and settling time of 3.5 secs, 1.7 secs and 0.5 secs, for the low, medium and high magnetic anchoring forces, respectively. The maximum transient error, when disturbed by a force approximately 4 times larger than required in the most demanding abdominal surgery was found to be 1mm in displacement using a high magnetic anchoring force. Significant attenuation of the mechanical disturbances (gain of 0.1 or less), due to the high stiffness and damping of the abdominal wall, was observed from 100rad/s in the frequency response. If a robotic manipulator was attached to the anchoring device, the typical operating frequency of a manipulator movements (0.1-1Hz) would still produce unattenuated disturbances. It is expected that some error compensation through suitable control strategies is required. These outcomes establish the basis for the controller design and the design specification of the active magnetically-anchored surgical devices, i.e. the speed of the surgical manipulator and the configuration of the anchoring set, such that approximately linear tissue model can be achieved with minimal disturbance impact onto the abdominal wall tissue.     

\end{abstract}


\begin{multicols}{2}

\section{INTRODUCTION}
\label{sec:intro}

In recent years, there have been minimally invasive surgical instruments that require a component of the device to be anchored to the inside of the abdominal wall during surgical procedures \cite{Zeltser_MAGS, Best_MAGS}. This approach enables the surgical component to be completely inserted into the abdominal cavity through a small incision and allows full mobility of the instrument along the wall of the insufflated abdominal cavity. The internal device, generally consists of a surgical instrument (e.g. surgical camera \cite{Hang_ACRA2015} or retractor \cite{Garbin_retractor}), is anchored to the inside of the abdominal wall through a magnet that is attracted to another magnet on the external side of the abdomen. In general, the anchoring location of the device can be manipulated by moving the external magnet to the desired location on the abdominal wall, which will result in the internal device being pulled to the same location on the inside of the abdomen. The general class of such instrument has been labeled as magnetically anchored and guided system (MAGS) \cite{Cadeddu_MAGS,Park_MAGS} and is illustrated in Figure \ref{fig:MAGS_illustration}. \\[3pt]

\begin{figurehere}
\centering
\includegraphics[width=0.43\textwidth]{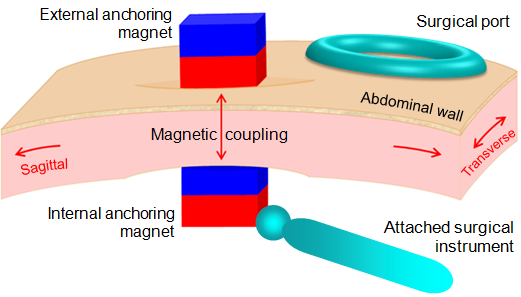}
\caption{Illustration of the MAGS device anchored onto the abdominal wall. The internal anchoring magnet with its attached surgical instrument (such as camera or tissue retractor) is inserted into the abdominal cavity through the surgical port at the incision.}
\label{fig:MAGS_illustration}
\end{figurehere}

A MAGS platform carries with it a surgical instrument that can be either passive or active (i.e. actuated). The passive devices are surgical instruments that only require anchoring in place during the surgical procedure. They are not actuated hence do not produce motion.  An example is a magnetic clip for tissue retraction, used to lift and then hold the tissue in place, as reported in \cite{Dominguez_MAGS_clip}. As the device is stationary during the surgical procedure, it does not produce any mechanical excitations to the abdominal wall.  The dynamics of the abdominal wall therefore does not affect the performance of the anchoring platform significantly. 

On the other hand, internal surgical devices that produce motions for surgical tasks manipulations within the abdominal cavity are categorised as active surgical devices. These anchored devices (e.g. robotic camera \cite{Hang_ACRA2015} and tissue retractor\cite{Garbin_retractor}) have  degrees-of-freedom (DOFs) actuated by either internally embedded onboard DC micromotors \cite{Rivas_MAGS_DC,Tortora_MAGS_DC} or through localised magnetic fields, known as localised magnetic actuation (LMA) \cite{DiNatali_LMA,Mohammadi_JMD,Mohammadi_TMECH,Leong_IROS2017}. These actuations result in interaction forces on the MAGS platform and excite the dynamics of the abdominal wall. In many cases, the abdominal wall is initially assumed rigid and the effect of the excitation caused by the actuation of the system is assumed insignificant. As the abdominal wall tissue exhibits the viscoelastic nature of a soft tissue, it is physically non-rigid.  As more challenging requirements in minimally invasive surgeries (MIS) pushed for more capable actuation to be realised for MAGS platforms in recent years \cite{Leong_RBME2016}, the effect of the actuation of the surgical instruments attached to the MAGS platform became more significant. This poses more challenges to the assumption mentioned especially in the aspects of positioning and accuracy, hence prompting for abdominal tissue dynamics to be taken into considerations in MAGS operation.

The properties and behaviors of the abdominal wall tissue during magnetic anchoring have yet to be investigated or taken into consideration for magnetically-anchored surgical methods. In more general cases, research on the abdominal wall tissue dynamics have been found in the area of surgical planning for MIS. The studies focused on the modelling of the abdominal insufflation \cite{Nimura_insuff} and the effects of insufflation onto the mechanical properties of abdominal wall tissue \cite{Song_stress_strain}. These studies have concentrated on the dynamics of the abdominal wall in the tangential direction of insufflation as it stretches. They do not address the dynamics of the wall in the direction of the magnetic anchoring, which is in the normal direction to the abdominal wall. 

To address this gap in the literature, this paper investigates the behaviour of the abdominal wall when subjected to magnetic anchoring and external disturbances in the normal direction as well as establishes a general numerical model of the abdominal wall tissue in the insufflation and magnetic anchoring conditions. The effect of the abdominal wall tissue onto the magnetically anchored surgical procedure is commonly ignored due to the assumptions that the tissue would experience high stiffness when insufflated and ``clamped" by anchoring magnets. Nonetheless, there are often queries on the extent of external disturbances from the actuation of active surgical devices affecting the abdominal wall tissue, and in turn creating undesired impact on surgical manipulations, such as movement and position inaccuracies. This question is addressed in this study by obtaining a numerical model of the abdominal wall tissue with emulated insufflation and magnetic anchoring. The model that closely approximates the responses of the tissue is then used to analyse the extent of the external disturbance caused by the motions and actuation forces due to active surgical instrument attached to the magnetic anchoring platform.

The remainder of the paper is structured in the following manner. Section \ref{sec:bg_abd_wall_model} discusses the background of the study related to the insufflated abdominal wall tissue and the implementation of existing findings in setting up realistic tissue modelling formulation for the tissue under the condition of magnetic anchoring and actuation. Section \ref{sec:mod_mtd} described the measurement procedures performed for the modelling of the abdominal wall tissue. The experimental results and the identification of the numerical model are presented in Section \ref{sec:results} along with an elaborated discussion in Section \ref{sec:discussion} on the obtained model and its dynamic behaviour when exerted by external loads. The findings in this study are then concluded in Section \ref{sec:conlusion}.



\section{Background on Insufflated Abdominal Wall}
\label{sec:bg_abd_wall_model}

During abdominal insufflation, the carbon dioxide (CO$_2$) gas is insufflated into the abdominal cavity creating a workspace within the abdomen for surgical procedures. The pressure due the insufflation lifts the abdominal wall tissue and causes the tissue to be stretched tangentially in the sagittal and transverse directions, as illustrated in Figure \ref{fig:Abd_wall_plane_insufflation}. Information on the abdominal wall during insufflation is essential to help the understanding of the tissue behaviour during magnetic anchoring and actuation on top of abdominal insufflation. Though investigations on the properties and behaviours of the abdominal wall tissue during magnetic anchoring and actuation have yet been performed, there are a couple of studies on the modelling of the tissue for laparoscopic surgical planning related to the abdominal workspace and tissue behaviour upon insufflation. \\[3pt]

\begin{figurehere}
\centering
\includegraphics[width=0.44\textwidth]{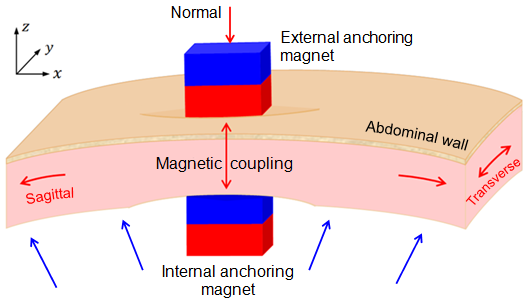}
\caption{Magnetic coupling on the abdominal wall in the normal (z-axis) direction to the tissue, with strains on the transverse and sagittal planes, due to abdominal insufflation.}
\label{fig:Abd_wall_plane_insufflation}
\end{figurehere}

\begin{figurehere}
\centering
\includegraphics[width=0.45\textwidth]{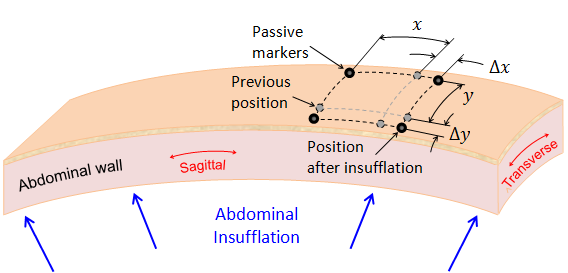}
\caption{In the study performed by \textit{Song et. al.}, passive markers are use to aid the measurement of the strains experienced by the abdominal wall tissue during insufflation, with $x$ and $y$ being the original distance between two markers in the sagittal and transverse directions respectively, before the insufflation. $\Delta x$ and $\Delta y$ depict the displacements of the markers from the original positions after the insufflation.}
\label{fig:Song_illustration}
\end{figurehere}

One such study was performed by \cite{Nimura_insuff} in which the abdominal insufflation is modelled as a mass-spring-damper system using the Finite Element Method (FEM) to describe the viscoelasticity nature of the CO$_2$ gas-insufflated abdominal cavity. The model parameters of the mass-spring-damper system identified are only based on the insufflation volume within the abdominal cavity and therefore, do not directly relate to the dynamics of the abdominal wall tissue. The authors did remark that the model would be more accurate if a mass-spring-damper model of the abdominal wall wass included as well. A more relevant study was performed in \cite{Song_stress_strain} with the analysis of the stress/strain relationship of the abdominal wall during insufflation. This study was performed on the abdomen of patients using passive markers and photogrammetry (as illustrated in Fig. \ref{fig:Song_illustration}). When the abdominal wall is lifted by the insufflation pressure, the markers on the abdomen experience some amount of displacement. Using the information of these displacements and the insufflation pressure, the mathematical model of the abdominal wall was analysed and developed in a FEM environment. It was observed that the abdominal wall experienced a strain of $5\%$ and $10\%$ in the transverse and sagittal planes, respectively, for the same stress it was subjected to by the insufflation. The findings reported on the behaviour of the tissue during insufflation do not involve the tissue in magnetically anchored and actuated conditions. It does however still provide an established starting point for this study in this paper by replicating the condition on a porcine abdominal tissue.

In the context of this paper, when the abdominal wall tissue experiences magnetic anchoring, it is subjected to the anchoring force in the normal direction on top of the tangential strains due to the insufflation pressure as reported in \cite{Song_stress_strain}. Hence, the effect of insufflation onto the abdominal wall was emulated in this study to closely simulated the abdominal surgical setting.


\section{Methodology for Tissue Dynamics Response Measurements}
\label{sec:mod_mtd}

In order to emulate the behaviour of the insufflated abdominal wall during magnetic anchoring, the parameter settings by \cite{Song_stress_strain} as discuss in Section \ref{sec:bg_abd_wall_model} are employed as the basis of the tissue measurement setup described in this section. The experimental data obtained will be used to approximate a general model of the abdominal wall tissue under insufflation and anchoring, which will be used for the analysis of the anchored tissue behaviour in response of external excitations, as discussed in Section \ref{sec:discussion}. This proposed model can also be incorporated into the electromechanical model of the LMA systems \cite{Mohammadi_TMECH, Leong_IROS2017,Leong_ACRA2016} for a complete system model for further implementation, e.g. controller design. It is important to note that even though the abdominal wall tissue consist of different anatomical layers, in this analysis, the abdominal wall tissue is assumed as a single mass with uniform properties ``clamped" or anchored by the external and internal units of the MAGS device. 

\subsection{Experimental Setup}
\label{subsec:exp_setup}

A benchtop experimental setup for the tissue measurement was constructed, as shown in Figure \ref{fig:Tissue_exp_setup}. The experiment setup uses a porcine model (i.e. pork belly) given the anatomical similarities between the human and porcine tissues \cite{Simon_pig}. The porcine belly specimens are trimmed to a specific dimension (350mm $\times$ 250mm) to fit into the experimental rig. These tissue specimens have an average thickness of 35mm (assuming uniform thickness across areas of measurements), which is within the range of the average human abdominal wall thickness (i.e. 20mm for averagely-built patients to 80mm for obese patients) \cite{Milad_thickness}. The measurement setup has to simulate the insufflated condition of the abdominal wall tissue in which the tissue experiences strains in the tangential directions. Hence, the tissue specimen was stretched corresponding to the length in which a strain of $10\%$ representing the strain due to insufflation observed in \cite{Song_stress_strain}. The selection of this parameter for the measurement setup is further discussed in Section \ref{subsec:set_insuff}. The skin side of the porcine belly specimen was placed on the tissue platform, exposing the inner side of the tissue for measurements as to provide realistic emulation of a surgical scenario, where the inner tissue is directly in contact with the internal surgical device. \\[3pt]

\begin{figurehere}
\centering
 \includegraphics[width=0.47\textwidth]{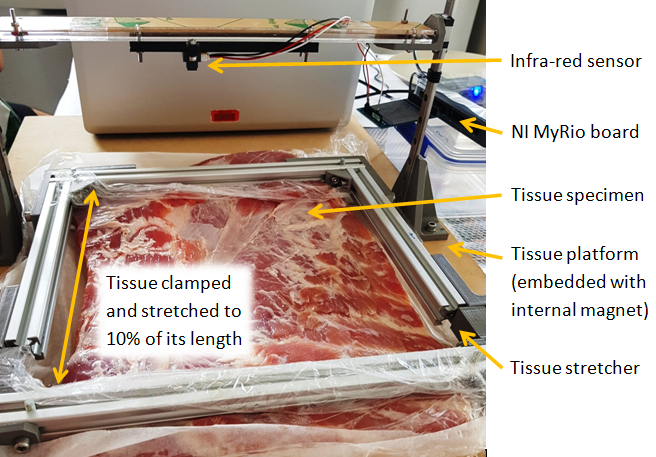}
\caption{Experimental setup for the simulated abdominal wall tissue measurements}
\label{fig:Tissue_exp_setup}
\end{figurehere}

\begin{figurehere}
\centering
\includegraphics[width=0.38\textwidth]{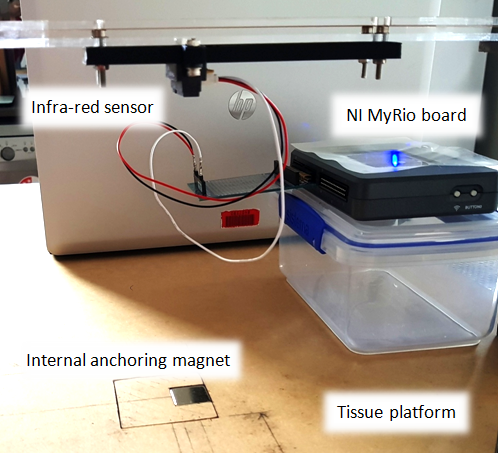}
\caption{Experimental platform without tissue specimen, revealing the internal anchoring magnet embedded in the tissue platform.}
\label{fig:IR_setup_int_anc}
\end{figurehere}

A pair of anchoring magnets are employed on either sides of the tissue in additional to the stretching. This represents the ``clamping" or anchoring by the MAGS device on the insufflated abdominal wall (see Fig. \ref{fig:Tissue_exp_setup_anc}). To provide a range of magnetic anchoring forces, three sets of anchoring magnets are selected for the study through coupling or anchoring force measurements using possible combinations of magnets from an available collection (i.e. dimensions of 10x10x10 mm, 20x20x12 mm, 38x38x12mm, and 50x50x12mm). The attraction force between the magnets is measured with incremental intermagnetic distance. The selection of the anchoring magnet sets based on the anchoring force measurements will be further discussed in Section \ref{subsubsec:anc_force}. \\[3pt]

\begin{figurehere}
\centering
\includegraphics[width=0.48\textwidth]{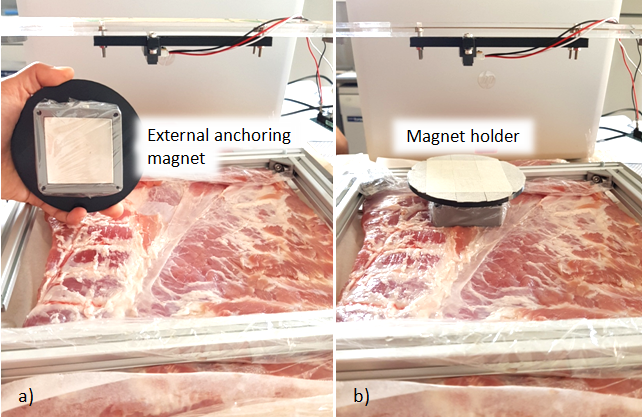}
\caption{Anchoring magnets are used provide an emulation of the tissue in the magnetically anchored condition, a) external anchoring magnet in a magnet holder, and b) external anchoring magnet coupled with an internal anchoring magnet which is embedded in the tissue platform right on the other side of the tissue specimen.}
\label{fig:Tissue_exp_setup_anc}
\end{figurehere}

The internal anchoring magnet from each set is embedded in the tissue platform so that the tissue specimen lays flat onto the platform during measurement (see Fig. \ref{fig:IR_setup_int_anc}). The motion response of the tissue during the magnetic anchoring is obtained by measuring the deformation or displacement of the tissue. These measurements are performed using a Infra-red (IR) sensor (0A41SKF54) placed above the tissue platform. The readings of the IR sensor against the elevated distance from the tissue platform are recorded to form a calibration curve to tissue displacement calibration. The procedure in generating the calibration curve to obtain the tissue displacements will be further elaborated in Section \ref{subsubsec:calib}. This calibration enables the approximation of the displacement the tissue experiences during magnetic anchoring as well as dynamic response tests.

\subsection{Parameter Settings for Emulating Insufflation}
\label{subsec:set_insuff}

To emulate the stretching of the abdominal wall tissue due to insufflation during a surgery, the stress/strain information presented in Figure 4 in \cite{Song_stress_strain} is used. At the given maximum insufflation pressure of 2.5 kPa, abdominal wall tissue experienced a strain of $5\%$ in the transverse direction and $10\%$ in the sagittal direction compared to the original size of the tissue. In our study, a strain of $10\%$ is applied to the porcine specimen in the measurement setup by stretching it to $10\%$ of its relaxed length to simulate the maximum insufflation pressure experienced by the abdominal wall tissue. 

\subsection{Calibration Process}
\label{subsec:calib_proc}

To obtain the responses of the tissue using the experimental setup described, displacement measurement and anchoring force calibration steps are executed prior to the experiments. These calibration steps are described in \ref{subsubsec:calib} and \ref{subsubsec:anc_force}, respectively.

\subsubsection{Calibration of Tissue Displacement Readings}
\label{subsubsec:calib}

The displacement of the tissue is determined by mapping the IR sensor readings to the respective calibration curve according to the anchoring magnets set, as stated in \ref{sec:mod_mtd}A. The calibration curves are obtained by recording the IR sensor readings as the height from the tissue platform gradually increases using measurement jigs, each with 5mm thickness, up to a minimum distance of 40mm between the IR sensor and the jigs, as illustrated in Figure \ref{fig:illus_jig}. The IR sensor will be used to measure the distance from a known constant height to the surface of the tissue specimen. Figure \ref{fig:Calibration_curve} shows the curve of IR readings versus the distance from the tissue platform. The calibration curve depicts an exponentially decreasing response as the inter-magnetic distance (i.e. the thickness of the abdominal wall tissue) increases. \\[3pt]

\begin{figurehere}
\centering
\includegraphics[width=0.28\textwidth]{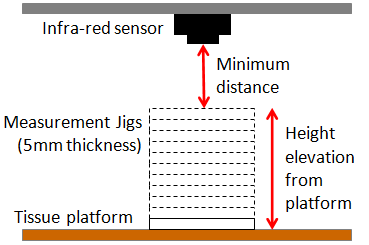}
\caption{Illustration of height measurements using the IR sensor and the measurement jigs.}
\label{fig:illus_jig}
\end{figurehere}

\begin{figurehere}
\centering
\includegraphics[width=0.45\textwidth]{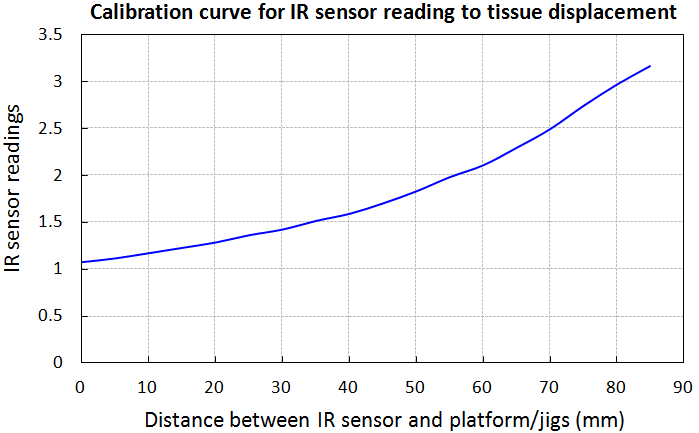}
\caption{Calibration curve for mapping of IR sensor readings to displacement or deformation of tissue.}
\label{fig:Calibration_curve}
\end{figurehere}

\subsubsection{Determining Anchoring Force across Tissue}
\label{subsubsec:anc_force}

The anchoring or attraction forces between all possible combinations of anchoring magnets from the available collection as mentioned in Section \ref{sec:mod_mtd} are measured with respect to their inter-magnetic distance using ATI Gamma force/torque sensor, as shown in Figure \ref{fig:anc_force_measurement}. A magnet of each possible combination is fixed onto the tissue platform while another is attached to the force/torque sensor (see Fig. \ref{fig:anc_force_measurement}). A linear actuator (Actuonix L12-50-100-12-P) is used to drive the force/torque sensor to produce a stable measurement of the coupling force as the force/torque sensor and the attached magnet moves upwards from the platform.

Similarly, this measurements are performed without the presence of the porcine specimen so that the magnets can be placed as close as possible to one another. However, the combinations of bigger magnets, particularly with the 38x38x12 mm and 50x50x12 mm magnets, have to be measured with bigger intermagnetic distance (e.g. 15mm or 20mm onwards) because of the extremely strong magnetic attraction. The anchoring force measurements are plotted in Figure \ref{fig:anc_force_plot}. Note that, due to size limitation of surgical ports (incision up to 25mm diameter) for insertion of surgical tools \cite{Soper_portsize}, only magnets of dimensions 10x10x10 mm and 20x20x12 mm can be taken as the internal anchoring magnets. Hence, the feasible combinations of anchoring magnets require internal magnet of either 10x10x10 mm or 20x20x12 mm dimensions. Note also that other magnetic attraction forces can be obtained using higher grades of rare earth magnets, e.g. NdFeb N40, N42, N45 \cite{MagnetMan}. \\[3pt]

\begin{figurehere}
\centering
\includegraphics[width=0.42\textwidth]{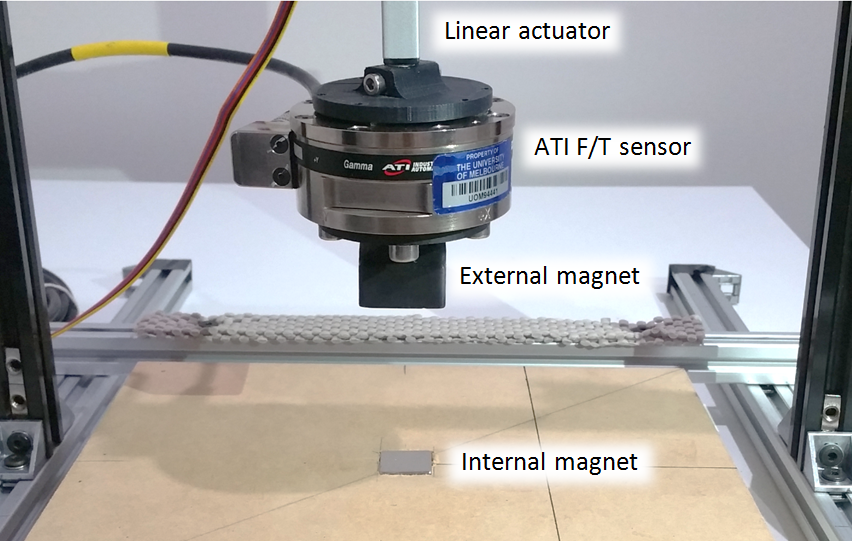}
\caption{Experimental setup for the measurement of coupling force between two anchoring permanent magnets}
\label{fig:anc_force_measurement}
\end{figurehere}

\begin{figurehere}
\centering
\includegraphics[width=0.47\textwidth]{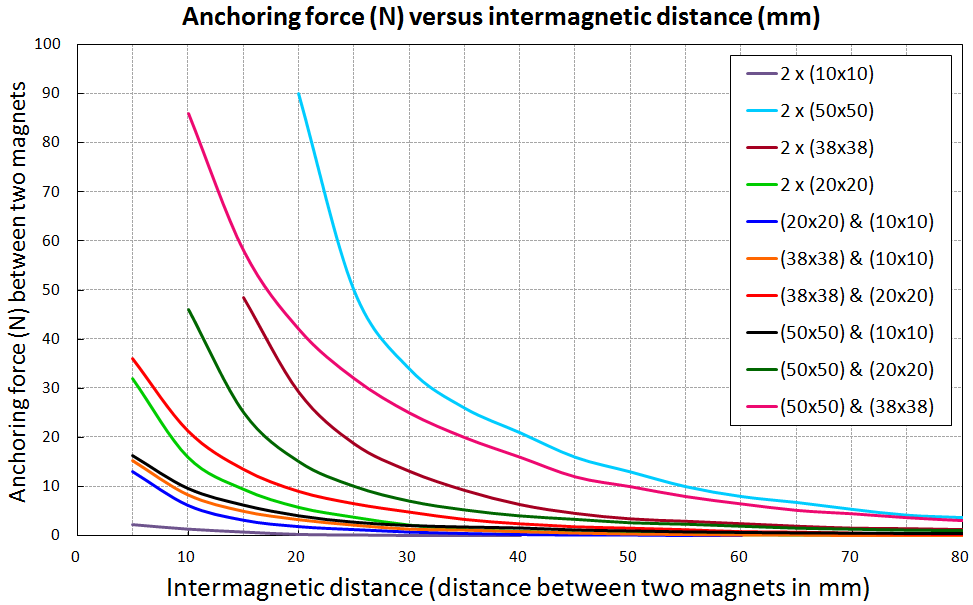}
\caption{Anchoring force measurements with respect to the intermagnetic distance, with possible combinations of magnets with different sizes. For instance, $2\times(10\times10)$ refers to both internal and external magnets of the same size (i.e. $10\times10\times10 mm$), and $(20\times20)\&(10\times10)$ refers to a combination of $20\times20\times12$ mm and $10\times10\times10$ mm magnets as the external and internal anchoring magnets, respectively.}
\label{fig:anc_force_plot}
\end{figurehere}

\subsubsection{Determining Thickness of the Tissue Specimen with and without Anchoring Magnets}
\label{subsubsec:anc_tissue_thickness}

The porcine specimen is placed on the platform below the IR sensor, and the IR readings at five locations around the tissue surface are recorded. The IR readings are mapped onto the calibration curve to obtain the thickness of the tissue. The average thickness across tissue is approximated to be 35mm.

To measure the resultant thickness of the tissue specimen with anchoring magnets, the largest possible choice for the internal anchoring magnet (i.e. 20x20x12 mm) and that for the external anchoring magnet (i.e. 50x50x12 mm) are then used to anchor onto the tissue specimen such that the maximum displacement the tissue experiences can be determined. The thickness of the tissue is measured using the IR sensor at five anchored positions across the specimen. Interpolating the IR sensor readings at this anchored position with the calibration curve (Fig. \ref{fig:Calibration_curve}), subtracting the height of the external anchoring magnet and its holder, gives the resultant tissue thickness of approximately 30mm in average. 

This shows that the original tissue specimen has been displaced approximately 5mm after maximum anchoring. At this stage, an equilibrium state due to the magnetic anchoring is achieved, as illustrated in Figure \ref{fig:cross_sec_abd_wall_coup}. Any perturbation or load onto the tissue on top of this anchored or equilibrium state, will produce an added displacement known as $\Delta z$ in this study. This equilibrium state aids the discussion on the investigation of the dynamic behaviour of the tissue under magnetic anchoring in Section \ref{sec:discussion}. \\[3pt]

\begin{figurehere}
\centering
\includegraphics[width=0.45\textwidth]{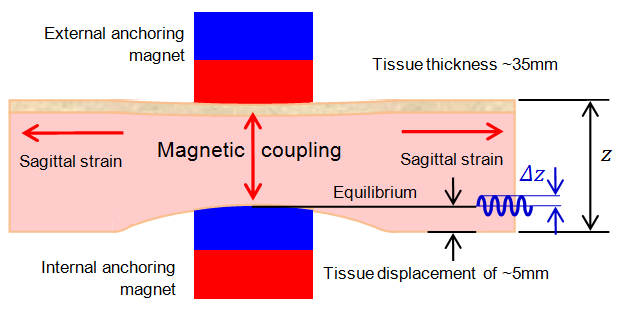}
\caption{Tissue experiencing the equilibrium state after magnetic anchoring. (NOTE: Tissue displacement not to scale, for visualisation purposes.)}
\label{fig:cross_sec_abd_wall_coup}
\end{figurehere}

\subsubsection{Determining the Selection of Anchoring Magnets Sets}
\label{subsubsec:anc_sets}

Table \ref{table:anchoring_sets} compiles the anchoring force measurements for the intermagnetic distance of 30mm. The feasible combinations of anchoring sets for the dynamic tissue analyses is selected based on the approximate range of anchoring forces required for common surgical tasks as listed in Table \ref{Table:SurgicalTasks}. The combinations of anchoring magnets with the highlighted and bold values of anchoring force shown in the table are then selected and listed below:
\begin{itemize}
\item Anchoring set 1: representing \textbf{low} anchoring force of 2N, with a combination of 50x50x12 mm magnet (external) and 10x10x10 mm magnet (internal) in this study,
\item Anchoring set 2: representing \textbf{medium} anchoring force of 4.7N, with a combination of 38x38x12 mm magnet (external) and 20x20x12 mm magnet (internal), and
\item Anchoring set 3: representing \textbf{high} anchoring force of 7N, with a combination of 50x50x12 mm magnet (external) and 20x20x12 mm magnet (internal).
\end{itemize}

These selections serve to provide a relevant range of anchoring forces versus the size of magnets for the investigation of the tissue dynamics as well as the validation of the proposed numerical model for this study in the following sections.

\begin{tablehere}
\tbl{Anchoring force (N) measurements of selection of anchoring magnets sets at a distance of 30mm. \label{table:anchoring_sets}} {
    \begin{tabular}{@{}m{2.8em} | c c c c@{}}
    \hline
	\multicolumn{1}{c}{Internal} & \multicolumn{4}{c}{External magnets} \\
	\multicolumn{1}{c}{magnets} & \multicolumn{4}{c}{(mm)} \\   [1ex]
    \multicolumn{1}{c}{(mm)} & \multicolumn{1}{c}{10x10x10} & \multicolumn{1}{c}{20x20x12} & \multicolumn{1}{c}{38x38x12} & \multicolumn{1}{c}{50x50x12} \\ [0.1ex]
    \hline \\ [-1ex]
        10x10x10 & 0.1 & 0.8 &  1.4 & \textbf{\hl{2}} \\ [2ex]
        20x20x12 & - & 2.2 & \textbf{\hl{4.7}} & \textbf{\hl{7}} \\  [2ex]
        38x38x12 & - & - & 13.1 & 25 \\  [2ex]
        50x50x12 & - & - & - & 34 \\  [2ex]
       \hline
    \end{tabular}}
\end{tablehere}

\begin{tablehere}
\tbl{Approximate force required in various surgical tasks. \label{Table:SurgicalTasks}}{
\begin{tabular}[width=0.45\textwidth]{@{}l r r@{}}
\hline
 Surgical tasks & Force & Ref  \\
 \hline 
 Liver and gall bladder retraction & 6.6 N &  \cite{Cadeddu_2008}  \\
    {Surgical camera} & {0.2 N} &  {\cite{Leong_RBME2016}}   \\
    {Surgical manipulation (e.g. pushing)} & {5 N}  & {\cite{Leong_RBME2016}}   \\
    {Soft tissue (e.g. liver penetration)} & {0.08 N} & {\cite{Leong_tissue}}  \\
    {Soft tissue (e.g. liver resection)} & {0.9 N}  & {\cite{Okamura_2004}}  \\
    {Suturing (on soft tissue, e.g. skin)} & {1.7 N}  & {\cite{Frick_2001}}   \\
\hline
      \end{tabular}}
\end{tablehere}

\subsection{Measurement of Tissue Mechanical Response to Mechanical Excitation}
\label{subsec:meas_tissue}

Loading and unloading tests are executed on the tissue specimen at its anchored and equilibrium condition using the three anchoring sets selected in Section \ref{subsubsec:anc_force} to obtain the tissue response. The tissue specimen is stretched to $10\%$ of its original lateral length to simulate the stress onto the tissue under insufflation based on \cite{Song_stress_strain}. Only the maximum direction of strain is assumed in this experiment due to the limitation of the setup and specimen.  

An internal anchoring magnet from the anchoring sets is placed below the tissue specimen and the corresponding external anchoring magnet is positioned right on top, across the porcine tissue to create an anchored condition on the tissue. The external anchoring magnet also allows the measurement of tissue displacement by the IR sensor placed right above when it moves with the surface of the tissue during the loading and unloading tests. A common load of 25N, which is 4 times the safety factor of the largest force required for the surgical tasks listed in table \ref{Table:SurgicalTasks} (i.e. max 6.6N), is used as an external excitation load on the tissue. Figure \ref{fig:load_unload_test} shows the measurement setup of the loading and unloading test with the 25N load placed on top of the external anchoring magnet holder. This load will remain on the anchoring magnet until the tissue reaches its steady state and then removed so the tissue creeps back to its relaxed condition. 

These tests are repeated on three locations around the porcine specimen and the responses are recorded to check the repeatability and accuracy of the measurement. The loading and relaxation responses are plotted and discussed in the results section (Sec. \ref{sec:results}) and are used to construct and identify the model of the tissue based on the general standard soft tissue model mentioned in Section \ref{sec:bg_abd_wall_model}. After obtaining the closest model to match the tissue responses, the mechanical parameters can then be identified as described in Section \ref{subsec:param_id}. \\[3pt]

\begin{figurehere}
\centering
\includegraphics[width=0.38\textwidth]{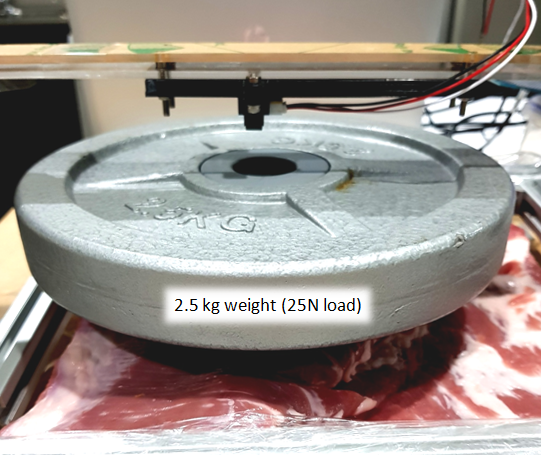}
\caption{Loading and unloading of force on the anchored tissue specimen using a 2.5kg weight to provide a 25N load.}
\label{fig:load_unload_test}
\end{figurehere}


\section{Experimental Results and System Identification}
\label{sec:results}

\begin{figure*}[!t]
\centering
\includegraphics[width=0.98\textwidth]{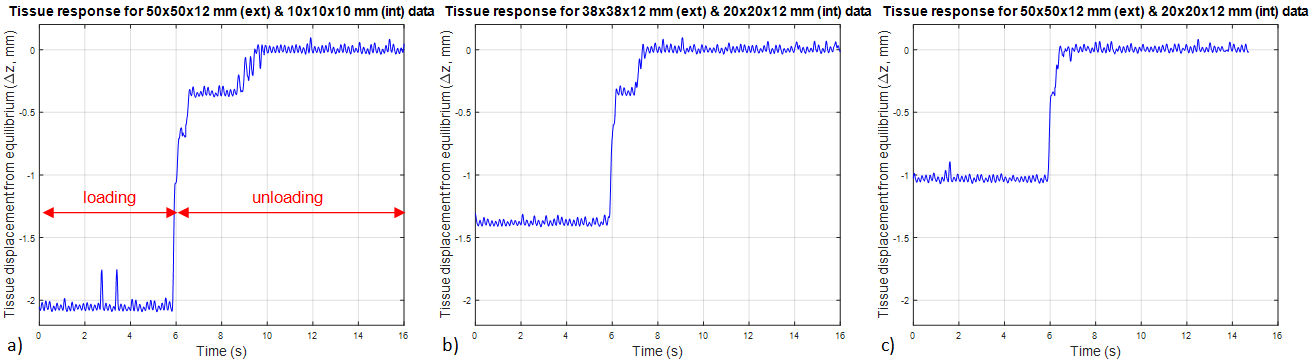}
\caption{The responses of the tissue specimen at its steady state during loading with 25N load and its creep responses when the 25N load is removed, i.e. on top of the tissue equilibrium with anchoring set consisting of, a) Anchoring set 1 (anchoring force approximately 2N), b) Anchoring set 2 (anchoring force approximately 4.7N), and c) Anchoring set 3 (anchoring force approximately 7N).}
\label{fig:load_unload_resp}
\end{figure*}

\begin{figure*}[!t]
\centering
\includegraphics[width=0.98\textwidth]{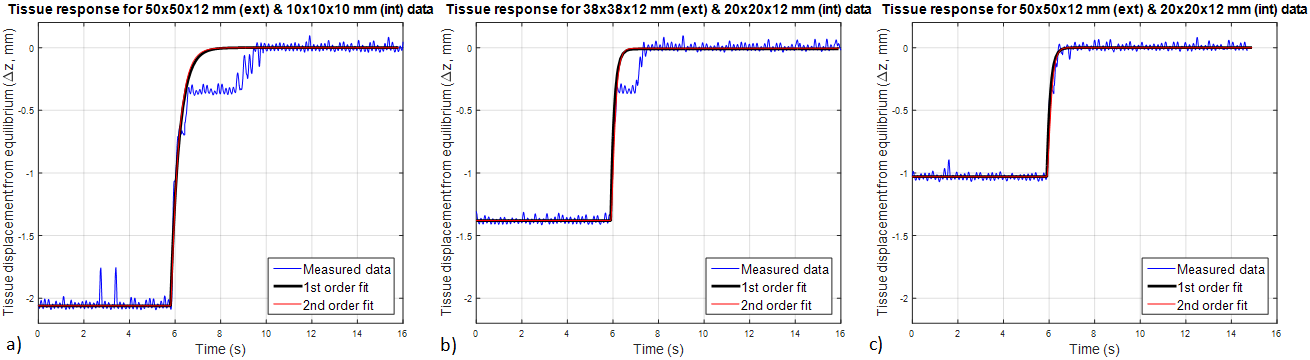}
\caption{The fitting of approximated standard first order (black plot) and second order (red plot) system responses onto the responses of the tissue specimen at its steady state during loading with 25N load and its creep responses when the 25N load is removed, i.e. on top of the tissue equilibrium with anchoring set consisting of, a) Anchoring set 1, b) Anchoring set 2, and c) Anchoring set 3.}
\label{fig:load_unload_resp_trials}
\end{figure*}

\begin{figure*}[!t]
\centering
\includegraphics[width=0.98\textwidth]{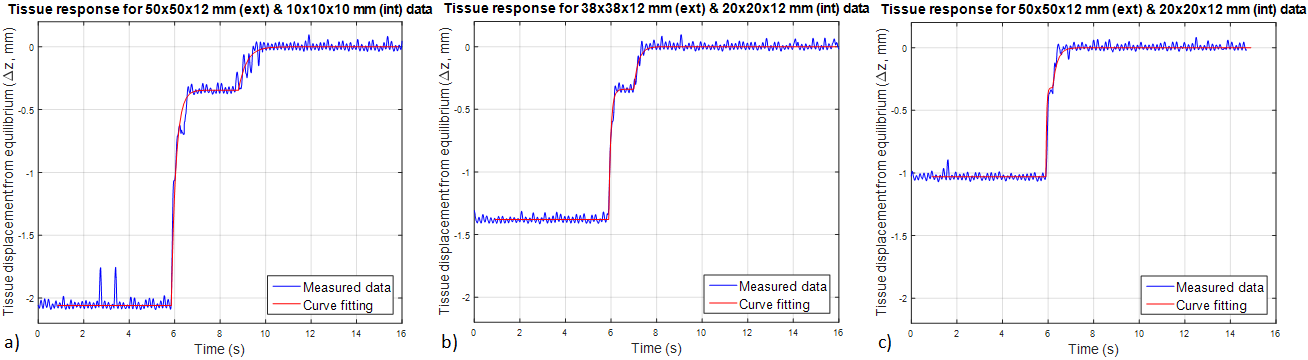}
\caption{The fitting of approximated general numerical model onto the responses of the tissue specimen at its steady state during loading with 25N load and its creep responses when the 25N load is removed, i.e. on top of the tissue equilibrium with anchoring set consisting of, a) Anchoring set 1, b) Anchoring set 2, and c) Anchoring set 3.}
\label{fig:load_unload_resp_fit}
\end{figure*}

This section presents the results of the tissue response experiments. There are three sets of experimental responses obtained using three selected anchoring magnets sets selected in \ref{subsubsec:anc_force}. These measurements are performed on three different locations of the porcine specimen which have approximately similar thickness and properties. These data are then plotted and approximated with the best fitted model with parameters identification as discussed in Section \ref{subsec:param_id}.

\subsection{Loading and Unloading Tests}
\label{subsec:result_load_unload}

Figure \ref{fig:load_unload_resp} shows a representative creep response of the tissue specimen for each of the three anchoring magnets sets (representing low, mid and high anchoring force) after the load has been removed. It is observed that the responses approximate a fast step response followed by another step response after a duration of delay which occurs at about similar displacement for all anchoring sets. This delay was observed to reduce as the anchoring force increases. Intuitively, the weaker the anchoring force, the higher the displacement the magnetically-anchored tissue experiences, hence the tissue with the smallest anchoring force (i.e. 2N, see Fig. \ref{fig:load_unload_resp}a) has the largest displacement due to the common load. Using the information obtained from the response, the parameters for a general numerical model that fits closely to all these responses can be estimated, as discussed in the next subsection.

\subsection{Proposed Numerical Model and Parameters Identification}
\label{subsec:param_id}

\begin{tablehere}
\tbl{Best fit parameters approximation for the tissue creep response across three data sets for all three anchoring sets. \label{table:creep_param}} {
\begin{tabular}{@{}lcccccc@{}}
\hline \\ [0.1ex]
Anchoring set $i$ & Data & $k_1$ & $\tau_1$ & $k_2$ & $\tau_2$ & $\theta$ \\ [1ex] 
\hline \\ [0.1ex] 
 & 1 & 1.715 & 0.2 & 0.345 & 0.35 & 2.95 \\ [1ex]
1 & 2 & 1.715 & 0.2 & 0.345 & 0.2 & 2.8 \\ [1ex]
 & 3 & 1.715 & 0.2 & 0.345 & 0.3 & 2.6 \\ [1ex]
\hline \\
Average &  & 1.715 & 0.2 & 0.345 & 0.283 & 2.783 \\ [1ex]
\hline
\hline \\ [0.1ex]
 & 1 & 1.04 & 0.09 & 0.34 & 0.2 & 1.1 \\ [1ex]
2 & 2 & 1.04 & 0.09 & 0.34 & 0.2 & 0.8  \\ [1ex]
 & 3 & 1.04 & 0.07 & 0.34 & 0.2 & 1.3 \\ [1ex]
\hline \\
Average &  & 1.04 & 0.083 & 0.34 & 0.2 & 1.067 \\ [1ex]
\hline
\hline \\ [0.1ex]
 & 1 & 0.71 & 0.035 & 0.32 & 0.2 & 0.3 \\ [1ex]
3 & 2 & 0.78 & 0.065 & 0.26 & 0.1 & 0.25 \\ [1ex]
 & 3 & 0.71 & 0.035 & 0.32 & 0.1 & 0.2 \\ [1ex]
\hline \\
Average &  & 0.733 & 0.45 & 0.3 & 0.133 & 0.25 \\ [1ex]
\hline
\hline 
\end{tabular}}
\end{tablehere}

From the results of the tissue unloading tests, the responses were first estimated using the standard the first and the second order system responses to observe the closeness of fit, as shown in Figure \ref{fig:load_unload_resp_trials}. These responses are tuned to best fit but could only closely approximate the initial step response of the tissue creep responses for all three anchoring sets. Although these responses approximate the tissue creep response with anchoring set 3 (i.e. high anchoring force) relatively well in comparison to that with anchoring sets 1 and 2 as observed in Fig. \ref{fig:load_unload_resp_trials}c, it is due to the smaller time delay resulting from the high anchoring force. The fitting approximation could not capture the observed time delay which is larger in the case of anchoring sets 1 and 2. In order to build a unified model for all three cases and closely accommodate the time delay observed, a summation of two single order step responses with a delay on the latter step response are used to approximate a general numerical model for the creep responses of these anchored tissue.

By computing $\tau$ and approximating the gains and time delays to find the best fit for the tissue creep responses, the following model transfer function can be written as:
\begin{equation}
{	
G_i(s) = \frac{k_{1_i}}{\tau_{1_i} s+1} + \frac{k_{2_i}}{\tau_{2_i} s+1}e^{-\theta_i s},
}
\label{eqn:1st_order}
\end{equation}

\noindent with $i=$1, 2, or 3 with respect to the anchoring sets. For each ``i'', $k_{1,i}$ and $k_{2,i}$ are the response gains, $\tau_{1,i}$ and $\tau_{2,i}$ are the time constants which is the time taken to reach 63.2$\%$ of the steady state values, and $\theta_i$ is the time delay in the response. 

The parameters for each data sets are presented in Table \ref{table:creep_param}, along with the average values of the parameters across the three response data for each anchoring sets. The average values are utilised to plot the transfer function responses to fit onto the tissue creep responses, as shown in Figure \ref{fig:load_unload_resp_fit}.

Therefore, the transfer function models estimated for the anchoring magnets sets incorporating the average parameters above are given as:

\begin{itemize}
\item Anchoring set 1:
\begin{equation}
{	
G_1(s) = \frac{1.715}{0.2s+1} + \frac{0.345}{0.283 s+1}e^{-2.783 s}.
}
\label{eqn:G1}
\end{equation}
\item Anchoring set 2:
\begin{equation}
{	
G_2(s) = \frac{1.04}{0.083 s+1} + \frac{0.34}{0.2 s+1}e^{-1.067 s}.
}
\label{eqn:G2}
\end{equation}
\item Anchoring set 3:
\begin{equation}
{	
G_3(s) = \frac{0.733}{0.45 s+1} + \frac{0.3}{0.133 s+1}e^{-0.25 s}.
}
\label{eqn:G3}
\end{equation}
\end{itemize}

\subsection{Tissue Response Analysis}
\label{subsec:analysis}

In order to further understand the behaviour of the tissue at its equilibrium state (i.e. anchored condition) towards external loads, the Bode plots of the transfer function using the models obtained above are utilised to show how the system will response for different input signals at different frequencies. The Bode plots for the tissue creep response model for all three anchoring sets are shown in Figure \ref{fig:Bode_creep_response}.

The plots show higher DC gain with the lowest the anchoring force (2N) while that close to zero with highest anchoring force (7N). The amplitudes of all three frequency responses decrease towards higher frequency operation. These results indicate that the dynamics of the system are only sensitive to low frequency noises with respect to the cutoff frequency of the models determined as 4.7 rad/s, 6.73 rad/s and 5.12 rad/s for  anchoring sets 1, 2 and 3 respectively. Nonetheless, the main interest of this analysis is the amount of attenuation the models offer at the operational speed of interest. 

\begin{figurehere}
\centering
\includegraphics[width=0.5\textwidth]{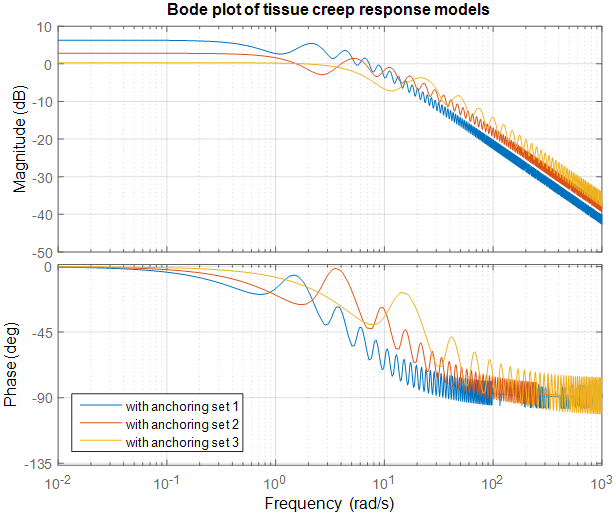}
\caption{Bode plot of transfer functions from obtained models.}
\label{fig:Bode_creep_response}
\end{figurehere}


\section{Discussions: Impact of the Dynamics on Magnetically Anchored Surgical Robotics Systems}
\label{sec:discussion}

Understanding the dynamics of the insufflated and magnetically-anchored abdominal wall tissue plays an important role in the operation of active devices, such as robotic surgical manipulator, anchored magnetically on the abdominal wall. Motion generated by the actuated devices would excite the mechanical dynamics of the abdominal wall and produce vibrations.  The following sub-sections discuss the results in the context of operating active magnetically anchored devices. 

\subsection{Maximum Displacement and Abdominal Wall Stiffness}
A load of 25N was applied in this paper to all anchored devices, simulating a disturbance force of 4 times of the maximum force reported in the literature encountered in abdominal surgeries (see Table \ref{Table:SurgicalTasks}). The maximum displacement the anchored tissue experiences is approximately 2mm when using anchoring set 1 with the lowest anchoring force (2N) (see Fig. \ref{fig:load_unload_resp}a). This displacement is reduced  to 1mm when anchoring set 3 with the highest anchoring force (7N) is used (see Fig. \ref{fig:load_unload_resp}c).  Note that the abdominal wall is insufflated (stretched) and further compressed by the anchoring (external and internal) magnets, thus the stiffness is already very high.  The results on Fig. \ref{fig:load_unload_resp}a
can be used to approximate the abdominal wall stiffness under such conditions (insufflated and magnetically anchored): 12.5 KN/m for the low anchoring force and 25KN/m for the high anchoring force.


\subsection{Steady State Response}
Fig. \ref{fig:load_unload_resp}a shows that the abdominal wall tissues upon unloading recovers completely to its displacement at the insufflated and magnetically anchored state, although a non-linear behaviour was observed where there is a component of fast spring-like recovery followed by another delayed damped response to the original displacement.  

\subsection{Transient Response}
Further analysis shows that the anchored tissue responses have considerably fast time constant at the beginning of its unloading response. However, there is a delay before it continues to its steady state displacement, resulting in a settling time that varies significantly with the anchoring force (i.e. approximately 3.5 secs, 1.7 secs and 0.5 secs for anchoring sets 1, 2 and 3 respectively). 
In this paper, we do not seek to explain the tissue biomechanics that produce this behaviour. The observed behaviour is identified for the purpose of designing and operating robotic mechanisms anchored to the anchoring system. It can also be observed from Figure 14 that the response of the abdominal wall tissue for the highest anchoring force can almost be approximated by a linear (step) response.

For active devices that are driven by external magnetic actuation signals, such as the LMA devices \cite{DiNatali_LMA,Mohammadi_JMD,Mohammadi_TMECH,Leong_IROS2017}, magnetic fields are generated and transmitted by external magnetic sources across the abdominal wall to actuate the permanent magnet rotor embedded in the internal device attached to the anchoring magnet. These sinusoidal actuating signals, transmitted using the magnetic field, are intended to interact with the permanent magnet rotor and produce actuation. Of course, there is the unintended effect that it also interacts with the anchoring magnet, creating periodic disturbance to the anchoring magnets. However, this effect is expected to be minimal as magnetic field decays exponentially with distance, and the anchoring magnets can be placed away from the center axis of the stators and rotor by design. Furthermore, due to high gear ratio typically used in such surgical manipulators, these actuation signals tend to be of high frequency, beyond the cutoff frequency of the abdominal wall, as further elaborated in the following section. 

\subsection{Frequency Response Analysis on Insufflated and Anchored Abdominal Wall}

To further understand the extent of potential disturbances introduced by mechanical dynamics of the anchored devices onto the insufflated and anchored abdominal wall, the Bode plots of the tissue responses are presented (see Fig. \ref{fig:Bode_creep_response}). In the case of the LMA devices, there are two potential sources of mechanical disturbances. The first is the magnetic actuation on the internal rotor. While the magnetic actuation produces the intended rotation of the rotor, it also produces unintended forces in the linear direction.  Relevant to our analysis here is the $z$ component of such force. However, due to the high gear ratio utilised in the typical internal surgical manipulators (e.g. gear ratio of 121.45:1 \cite{Hang_ACRA2015}), such actuation of the rotor is carried out at high frequencies beyond the cutoff frequencies of the abdominal wall, which would be significantly attenuated.   

The second component of mechanical disturbance is generated by the motion of the surgical manipulators themselves. This applies to not only LMA based devices but also to all magnetically anchored active surgical devices. The operating frequency of such surgical manipulators is commonly low in the range of 0.1 Hz to 1 Hz (i.e. 0.63 rad/s to 6.3 rad/s) \cite{Hu_rads, Terry_rads}. This component of disturbance will not be significantly attenuated.  Attention should be paid to this type of disturbance when designing magnetically anchored surgical devices. Such disturbances can be accommodated for by the controller design.

\subsection{Other General Observations}

It can be observed that the high anchoring force (produced by anchoring set 3) resulted in: higher abdominal wall stiffness, lower overall displacement due to disturbance forces, and much shorter delay in the unloading step response (creep) (Fig. \ref{fig:load_unload_resp}c), resulting in a close approximation of a linear step response (Fig. \ref{fig:load_unload_resp_fit}c).  It was not observed to significantly lower the cut-off frequency of the abdominal wall response. 

With this knowledge, the actuation requirements, such as the velocity of the internal device can be configured such that the disturbance onto the abdominal wall tissue during surgical manipulation is minimised. On the other hand, with a good numerical model, it is possible to validate if the proposed control law has the capability to cope with such disturbance. Moreover, this disturbance model might be directly used to facilitate the control design using the technique such as Internal Model Principle \cite{Hippe_IMP, Landau_IMP, Leong_RAL2018}.


\section{Conclusion}
\label{sec:conlusion}

The modelling of the abdominal wall tissue in the insufflated and anchored condition was carried out to improve the understanding of the tissue behaviour in the face of magnetically anchored surgical devices. With the attachment of active surgical devices onto the anchoring magnets, there is a high possibility that the actuation signals or generated motions would cause disturbance or vibrations onto the abdominal wall tissue, hence resulting in inaccurate positioning or manipulation when performing surgical tasks. Investigating abdominal wall tissue response in these conditions with an applied load of 4 times the maximum force typical of abdominal surgical tasks listed in this study allows the observation of the tissue behaviour with respect to a range of the anchoring forces. 

It was observed that the abdominal wall actually introduces damping to the magnetically anchored systems, resulting in a low-pass filtering effect to any mechanical excitation. The outcomes, summarised in Section 5, indicated that the mechanical movements of an attached surgical device are likely to be the main contributors to disturbances that are not significantly attenuated by the abdominal wall tissue dynamics. Nonetheless, the disturbances can be managed by device configuration and controller design.

The future work will focus on testing the robustness of the proposed disturbance rejection controller developed for the system in \cite{Leong_RAL2018} when the obtained numerical model of the insufflated and anchored tissue response is  incorporated into the dynamic model of the LMA system \cite{Leong_IROS2017}. Moreover, new disturbance rejection algorithms can be developed with the consideration of disturbance model identified with experimental validations.




\noindent\includegraphics[width=1in]{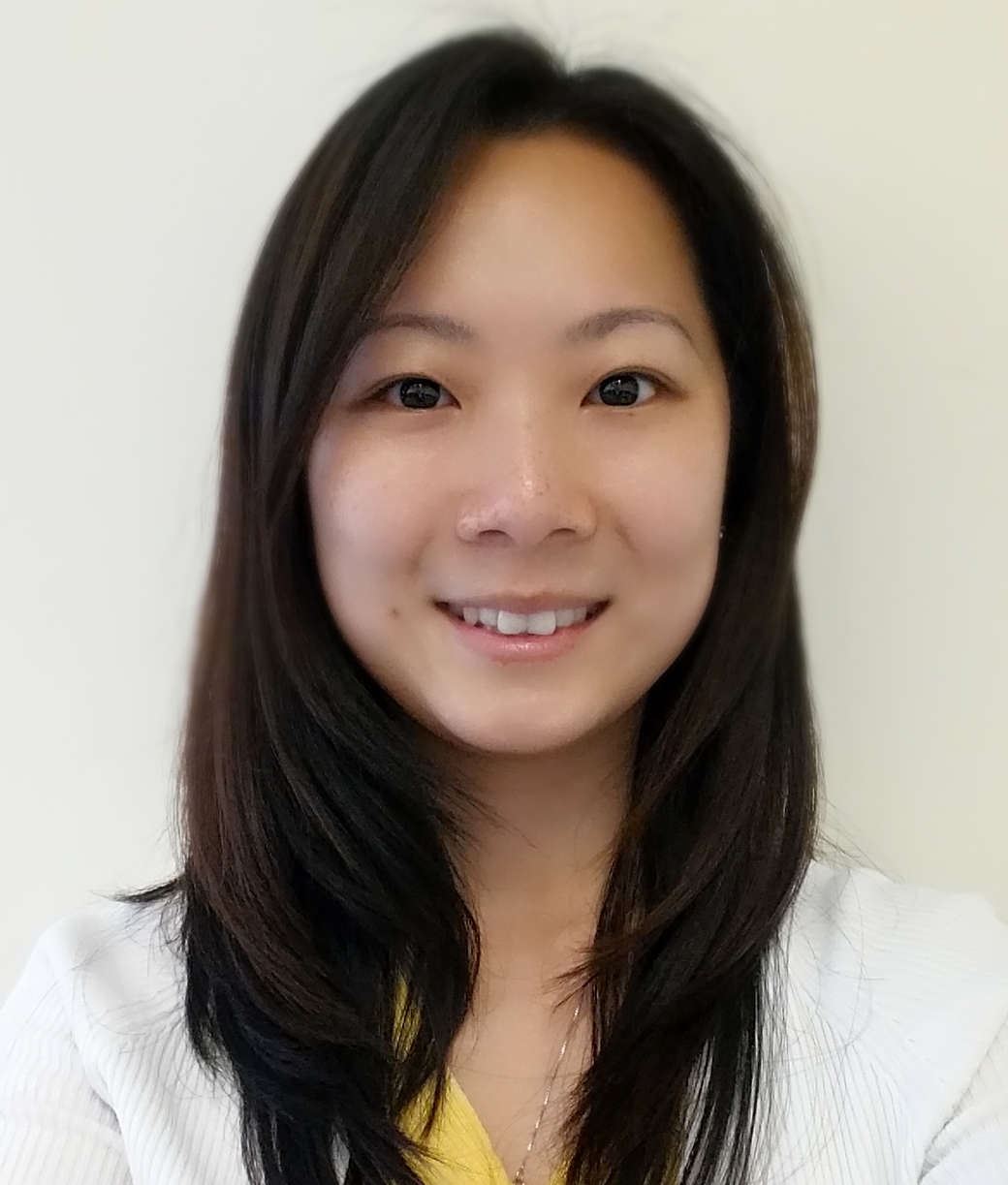}
{\bf Florence Leong} received the M.Eng. degree in biomechatronics and medical robotics from the National University of Singapore, Singapore. She is currently working toward the Ph.D. degree in medical robotics in the University of Melbourne, Melbourne, Australia. 

Her research interests include mechatronics and robotics applications in the area of medical and surgery. Her current research includes the development of a transabdominal local electromagnetic actuator for abdominal surgery. \\[12pt]

\noindent\includegraphics[width=1in]{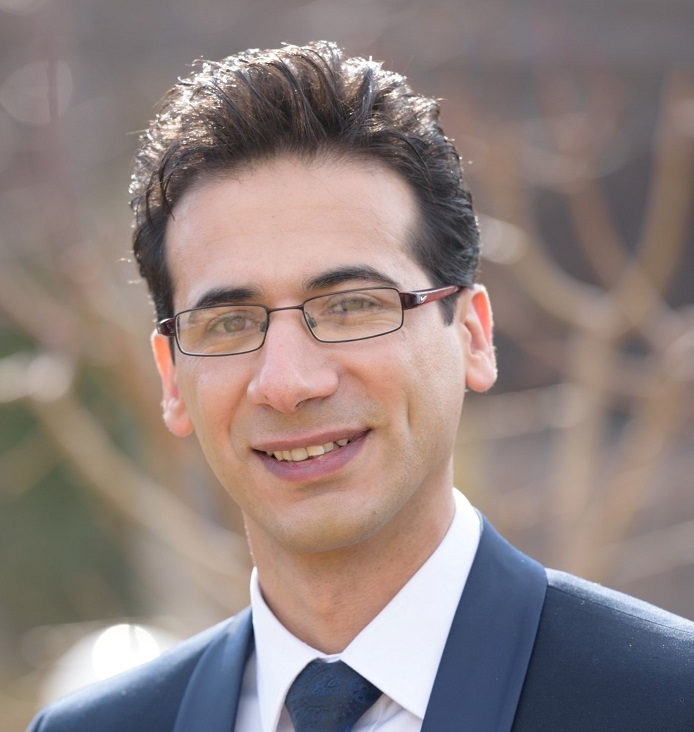}
{\bf Alireza Mohammadi} received the B.S. degree in mechanical engineering from the Iran University of Science and Technology, Tehran, Iran, in 2005, the M.Sc. degree in mechanical engineering from the Sharif University of Technology, Tehran, in 2008, and the Ph.D. degree from the University of Melbourne, Melbourne, Australia. 

He is currently a Postdoctoral Research Fellow
with the University of Melbourne, performing research on robotic surgery and control of mobile robots. His research interests include control, robotics, and biomedical engineering.\\[12pt]

\noindent\includegraphics[width=1in]{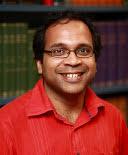}
{\bf Vijay Rajagopal} Dr Vijay Rajagopal received his undergraduate Bachelor of Engineering degree (1st Hons) in Engineering Science at the University of Auckalnd, NZ. He was awarded a PhD in Bioengineering from the Auckland Bioengineering Institute (ABI), University of Auckland, NZ in 2007. 

He conducted post-doctoral research at the Auckland Bioengineering Institute from 2007-2011 and the Singapore-MIT Alliance for Research and Technology from 2012-2013 before joining the University of Melbourne as a Senior Lecturer in Biomedical Engineering in 2014. His research expertise and interests span across cell and tissue remodeling, biomechanics and mechanobiology and computational physiology.  \\[12pt]

\noindent\includegraphics[width=0.8in]{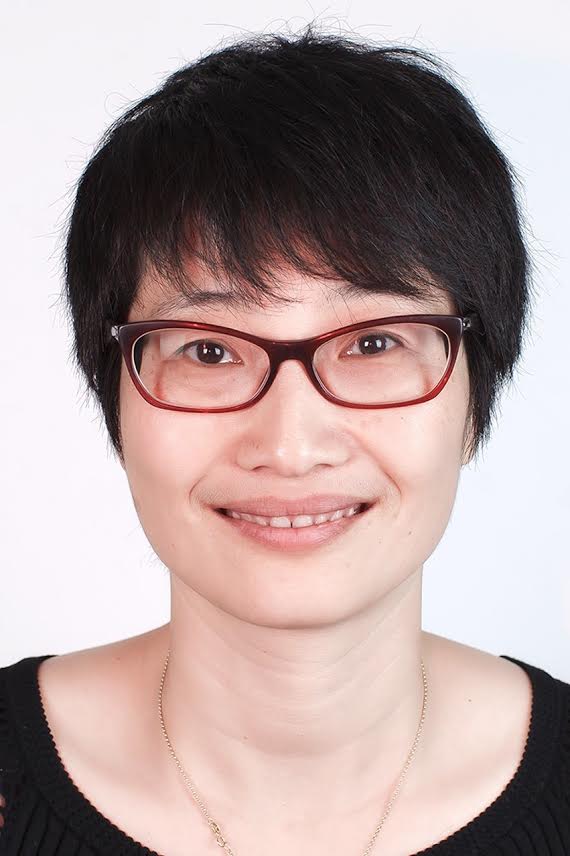}
{\bf Ying Tan} is an Associate Professor and Reader in the Department of Electrical and Electronic Engineering (DEEE) at The University of Melbourne, Australia. She received her Bachelor’s degree from Tianjin University, China, in 1995, and her PhD from the National University of Singapore in 2002. 

She joined McMaster University in 2002 as a postdoctoral fellow in the Department of Chemical Engineering. Since 2004, she has been with the University of Melbourne. She was awarded an Australian Postdoctoral Fellow (2006-2008) and a Future Fellow (2009-2013) by the Australian Research Council. Her research interests are in intelligent systems, nonlinear control systems, real time optimization, sampled-data distributed parameter systems and formation control. \\[12pt]

\noindent\includegraphics[width=1in]{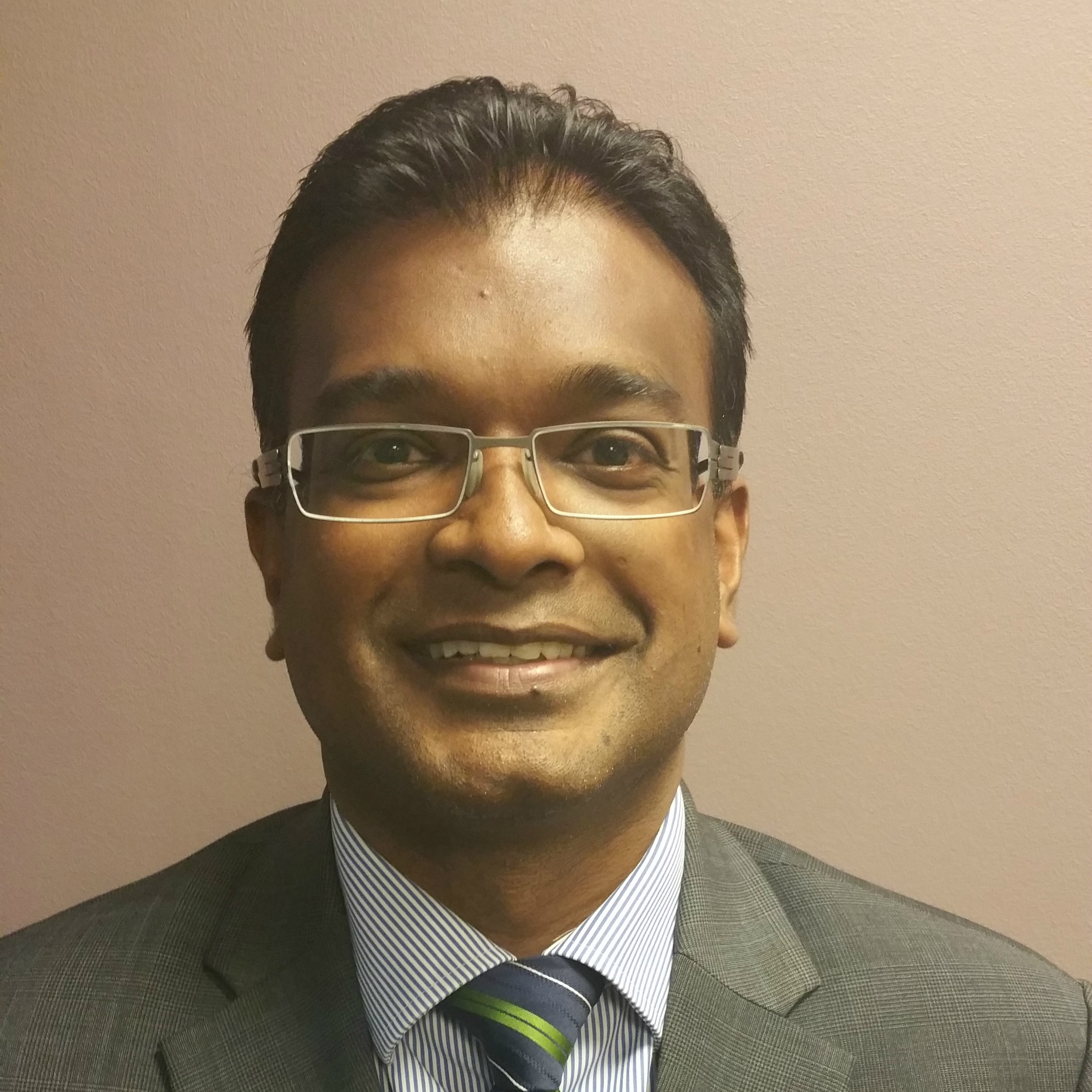}
{\bf Dhan Thiruchelvam} received the Bachelor of Medicine/Bachelor of Surgery (MBBS) degree from the Sydney UniversityMedical School, New
South Wales, Australia, and the Diploma degree in clinical education from the University of Edinburgh, Edinburgh, U.K., in 2011.

He is currently an Upper Gastrointestinal and Obesity Surgeon with the St. Vincents Hospital, Melbourne, Australia, and a Senior Lecturer for
the Department of Surgery, University of Melbourne, Melbourne. His research interests include the areas of Robotics Surgery and Surgical education. Dr. Thiruchelvam is a Fellow of the Royal Australasian College of Surgeons. \\[12pt]

\noindent\includegraphics[width=1in]{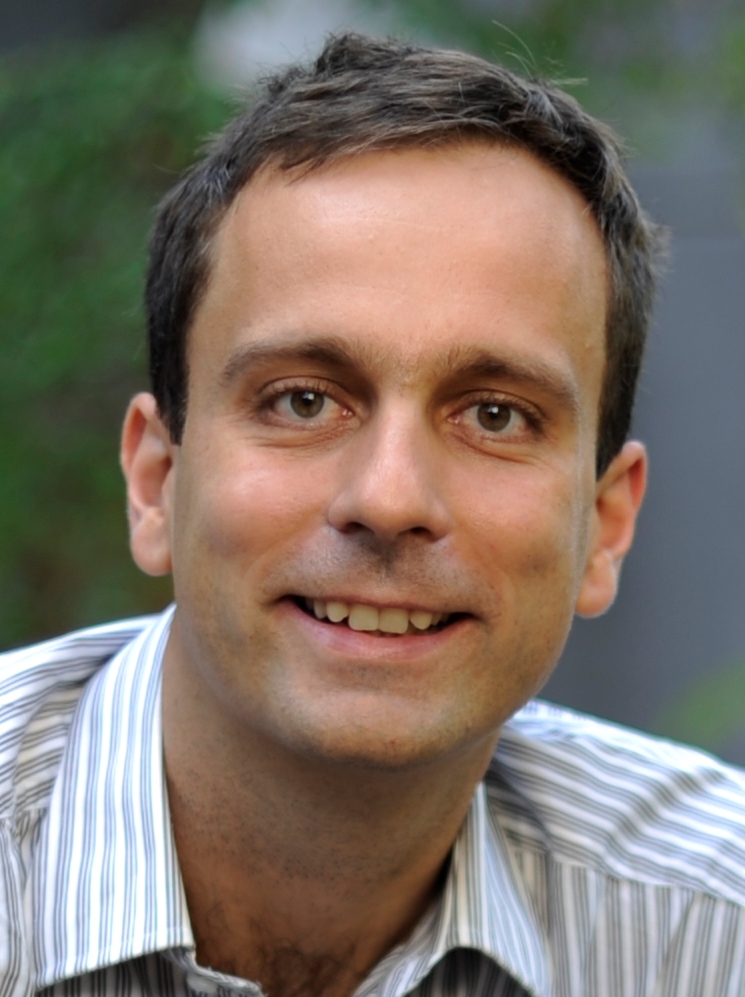}
{\bf Pietro Valdastri} (M’05–SM’13) received the Master (Hons.) degree in electronic engineering from the University of Pisa, Pisa, Italy, in 2002, and the Ph.D. degree in biomedical engineering from the Institute of BioRobotics of Scuola Superiore SantAnna (SSSA), Pisa, Italy.

He spent three years as Assistant Professor at the SSSA, and since 2011, he has been an Assistant Professor at the Department of Mechanical Engineering, Vanderbilt University, Nashville, TN, USA, where he founded the STORM Lab. He also holds secondary appointments in the Department of Electrical Engineering and in the Division of Gastroenterology, Hepatology and Nutrition, Vanderbilt University. He is actively involved in robotic endoscopy, robotic surgery, and design ofmagnetic mechanisms.

Dr. Valdastri received the NSF Career Award to study and design capsule robots for medical applications, in 2015. \\[12pt]

\noindent\includegraphics[width=1in]{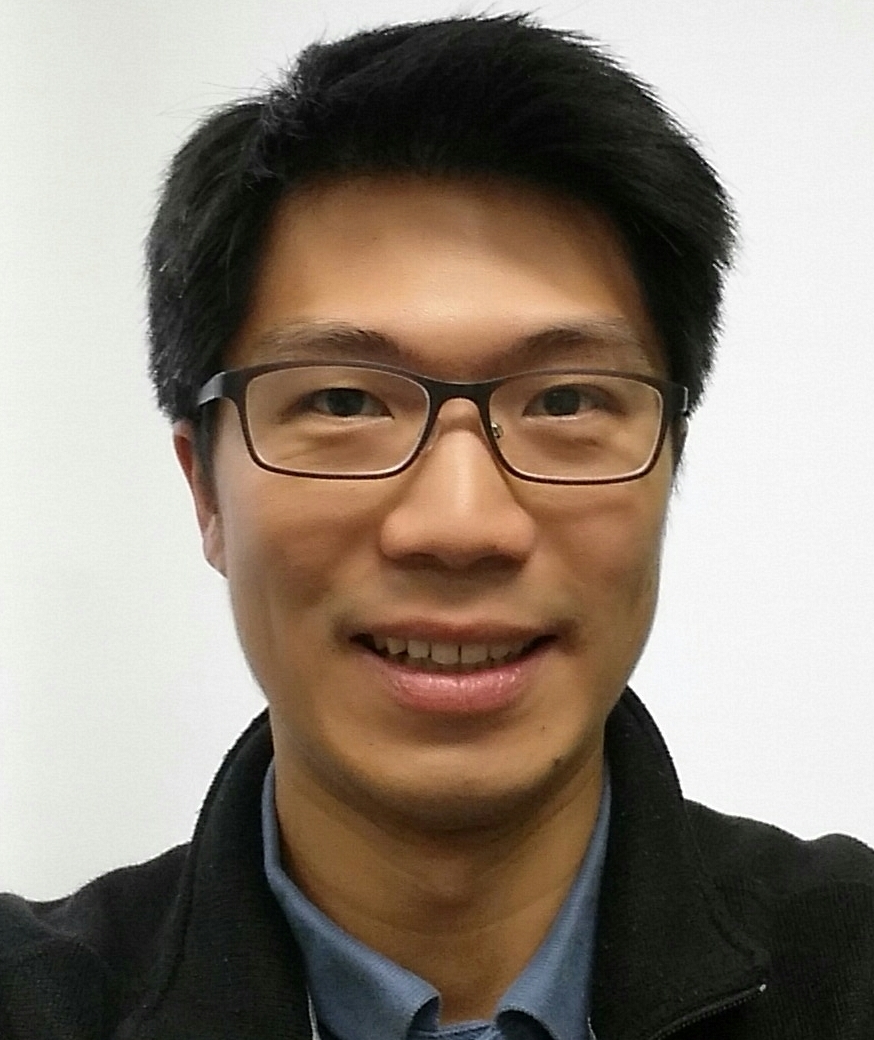}
{\bf Denny Oetomo} received the B. Eng. (Hons.) degree from the Australian National University, Canberra, Australia, in 1997, and the Ph.D. (mechanical engineering, robotics) degree from the National University of Singapore, Singapore, in 2004.

His research interests include the area of manipulation of system dynamics, with applications mainly in the areas of clinical / biomedical engineering. He was a Postdoctoral Fellow at Monash University (2004–2006), INRIA Sophia Antipolis (2006–2007), and joined the University of Melbourne, Melbourne, Australia, in 2008, where he is currently an Associate Professor. \\[12pt]

\end{multicols}
\end{document}